\documentclass{article}

\usepackage{PRIMEarxiv}

\usepackage[utf8]{inputenc} 
\usepackage[T1]{fontenc}    
\usepackage{hyperref}       
\usepackage{url}            
\usepackage{booktabs}       
\usepackage{amsfonts}       
\usepackage{nicefrac}       
\usepackage{microtype}      
\usepackage{lipsum}
\usepackage{fancyhdr}       
\usepackage{graphicx}       
\usepackage{caption}
\usepackage{subcaption}
\usepackage{amsmath}
\graphicspath{{media/}}     

\title{Accelerated Solutions of Coupled Phase Field Problems Using Generative Adversarial Networks
}

\author{
  Vir Karan \\
  Department of Metallurgical and Materials Engineering \\
  Indian Institute of Technology Madras \\
  Chennai \\
   \And
  A. Maruthi Indresh, Saswata Bhattacharya \\
  Department of Materials Science and Metallurgical Engineering \\
  Indian Institute of Technology Hyderabad \\
  Hyderabad \\
}

\begin{document}
\maketitle

\begin{abstract}
Multiphysics problems such as multicomponent diffusion, phase transformations in multiphase systems, alloy solidification involve numerical solution of a coupled system of nonlinear partial differential equations (PDEs). 
Numerical solutions of these PDEs using mesh-based methods require spatiotemporal discretization of these equations. Hence, the numerical solutions are often sensitive to discretization parameters and may have inaccuracies (resulting from grid-based approximations). Moreover, choice of finer mesh for higher accuracy make these methods computationally expensive. 
Neural network based PDE solvers are emerging as robust alternatives to conventional numerical methods because 
these use machine learnable structures that are grid-independent, fast and accurate. 
However, neural network based solvers require large amount of training data, thus affecting their generalizabilty and scalability. These concerns become more acute for coupled systems of time-dependent PDEs.
To address these issues, we develop a new neural network based framework that uses encoder-decoder based conditional Generative Adversarial Networks with \emph{ConvLSTM} layers to solve a system of Cahn-Hilliard equations. 
These equations 
govern  microstructural evolution of a ternary alloy undergoing spinodal decomposition when quenched inside a three-phase miscibility gap. We show that the trained models are mesh and scale-independent, thereby warranting application as effective neural operators.
\end{abstract}


\section{Introduction}
\label{sec:Intro}
Time-dependent partial differential equations (PDEs) are commonly encountered in various domains of science and engineering. 
Since most of these PDEs are nonlinear, analytical solutions do not exist. As a result, we need to develop efficient numerical methods to solve these problems. Numerical approaches such as finite difference/finite element/spectral methods involve discretization of  the domain under 
using a suitable mesh and finding accurate and efficient solvers. 
Accuracy of numerical solutions inversely depends on the coarseness of the mesh used while discretizing the domain, and is highly sensitive to the discretization parameters used. 
Although these techniques have been used extensively in the past, they are computationally expensive and generally do not scale well with the size of the mesh taken while discretizing the domain. 

Since the microstructure of a material provides a vital link between the atomic structure and the macroscopic properties of the material, tuning of the microstructure using different processing routes has been an effective tool to control the macroscopic behaviour of a material. A popular method to model microstructure evolution is the Phase-Field method, which involves solving the Cahn-Hilliard and the Allen-Cahn Equations to simulate the evolution of conserved and non-conserved parameters respectively. When applying this method, one often ends up with several coupled non-linear PDEs that have to be solved numerically. One such Phase-Field problem is modelling the evolution of the microstructure following the process of spontaneous phase decomposition in ternary systems i.e three phase systems, which is governed by two coupled Cahn-Hilliard equations, one for each independent concentration field, the solution of which is highly non-linear. 

Recent years have seen the rise of deep learning-based models that are capable of progressively extracting higher-level features from structures such as images, signals, graphs and text, and making accurate predictions from data directly in the domain of computer vision, speech recognition, text prediction, robotics etc. Over the last decade, deep neural networks have also found their way into solving partial differential equations \cite{Guo2016ConvolutionalNN, Lu2019DeepONetLN, Raissi2019PhysicsinformedNN, Zhu2018BayesianDC, wang2020physicsinformed}. Such networks can be split into two categories: solvers on discretized domains and solvers on continuous domains. Similar to conventional numerical solvers, the former networks work on discrete domains and tend to be mesh dependent. Such networks utilize convolution operations and are hence lightweight and scalable. The latter is capable of training on one mesh size and geometry, and making fairly accurate predictions for another mesh size or geometry, i.e. they are mesh independent. However, these tend to be slower to train and require large amounts of data. A recent advancement is the domain of neural operators \cite{Li2020NeuralOG, Li2020MultipoleGN, li2021fourier} that learn the mapping between infinite-dimensional function spaces and  hence, fall into the second category of solvers mentioned above. 

Another advancement in this field is the development of Physics Informed Neural Networks (PINNs)\cite{Raissi2019PhysicsinformedNN}. In addition to conventional loss functions, these networks add a residual of the governing PDE to the loss while training the networks. This allows for more accurate predictions, lower training data requirements and faster convergence while training. Although such networks appear promising, most real-world problems have data contaminated with noise thereby making obtaining the exact governing PDEs a challenging task. Furthermore, most works involving such deep learning-based methods have attempted to solve systems with only a single governing PDE, and these methods tend to fare poorly when dealing with coupled PDEs due to the additional non-linearities that are introduced into the system. Moreover, PDEs tend to further depend on additional parameters (physical properties, geometric characteristics, boundary conditions) that have to be obtained from data. Hence, there has been a lot of effort in recent years put into developing a deep learning-based method to solve coupled PDEs efficiently that doesn’t rely on the exact form of the governing PDEs, is scalable and requires a low amount of data to train. 

In the past, neural networks having a recurrent structure such as Long-Short Term Memory networks (LSTM) \cite{sak2014long} have been applied extensively when dealing with temporal data. A related development has been the Convolutional-LSTM Network \cite{Kim2017DeepRainCN}, which has been successfully applied to problems involving spatio-temporal data. Interestingly, several parallels can be drawn between LSTMs and Finite Elements and Finite Differences based methods, and an analysis of the same has also been done \cite{hu2020neural}, thereby warranting their use in the task of solving spatio-temporal PDEs.

Early attempts to solve phase-field problems using data-driven methods \cite{brough2017microstructure, fast2011new} attempted to learn the Green's function solution to PDEs by performing regression in the complex domain to learn the influence kernel in fourier space. This kernel was then used to predict the temporal evolution of the conserved phase-field order parameter in a binary system. Although this approach is highly efficient and can be applied to a multitude of problems involving a single PDE, it was found to be not as effective when dealing with coupled PDEs. PINNs \cite{wight2020solving} have shown their efficacy towards predicting the dynamics of a variety of systems governed by a single phase-field PDE by using adaptive sampling methods to efficiently capture the dynamics of moving interfaces. However, the approach presented utilized fully-connected neural networks, which are strictly dependent on the mesh used while discretizing the data used for training, and can require a large volume of data and computational time to train. Another recent notable work is the DeepXDE \cite{lu2021deepxde} library that used PINNs to solve differential equations. Although quite versatile, effective extension of the method used in this work to complex coupled PDEs is yet to be attempted. More recent attempts at solving time-dependent PDEs have used recurrent neural networks \cite{yang2020selfsupervised, de2021accelerating}, by posing the problem of solving spatio-temporal PDEs as a sequence-to-sequence translation problem. However, these still have shortcomings with regard to the accuracy and scalability of the trained models. Some works have used adversarial learning to complement the encoder-decoder based networks while solving PDEs such as the Navier-Stokes equation for incompressible fluid flow \cite{thavarajah2020fast} or the Cahn-Hilliard equation for phase segregation in binary alloys \cite{farimani2018deep}. These works have found that the addition of adversarial loss to the loss function improves the accuracy of the trained networks, especially in cases where there exist sharp gradients in the data that evolve in a highly non-linear fashion with time, which is especially true in the case of moving interfaces in most phase-field problems. Therefore, adding an adversarial component to the loss function while training is also a modification that can yield better results\cite{farimani2018deep}. We would also like to point out that previous works in this field using similar frameworks have either not used a recurrent structure to model the temporal dependencies in the data or not used adversarial training. Furthermore, to the best of the author's knowledge, there has not been an attempt at solving complex coupled phase-field problems with more than one independent field using neural network based methods in literature, marking this work to be the first of this kind.

Through this work, we propose a Conditional Generative Adversarial Network with an Encoder-Decoder Generator architecture having a recurrent structure to solve coupled spatio-temporal PDEs. We apply this model towards learning the dynamics behind the evolution of a ternary alloy undergoing phase decomposition and show that the trained model is generalizable to new initial conditions, is efficient and highly scalable, and capable of training on data from a coarse grid to make accurate predictions on a grid of larger geometry. Furthermore, we also show that the trained model is mesh-invariant, by showing the model trained on a square mesh shows similar performance on a rectangular mesh as well. Considering the success of our model on predicting microstructure evolution governed by the highly non-linear coupled Cahn-Hilliard equations, we hypothesize that the proposed model can be applied towards predicting other physical phenomena that are governed by non-linear PDEs as well. An outline of this work is as follows: Section 2 formulates the Phase-Field problem that we aim to solve and provides details of the network architecture that we propose, the training datasets and algorithms. Section 3 deals with analysing the predictions made by the trained model, by firstly analysing the ability of the model to learn the operator that governs the dynamics of the system, followed by an emphasis on the accuracy, scalability and computational complexity of the trained model. We also compare the performance of the model for differing temporal spacing between subsequent data-points, which is a crucial aspect of selecting the data to feed to the Neural-PDE solvers that is often not discussed. We conclude with Section 4 and shed light on possible directions for future work.

\section{Methods}
\label{sec:Methods}
\subsection{Problem Formulation}
\label{Problem Formulation}

As a function of location, i.e., the spatial coordinates $x$ and $y$, and time $t$, let $c_i(x, y, t)$ represent the mole fraction of the "i"th component of a multi-component material. We consider three components A, B, and C for a ternary system such that: 
\begin{equation}
\label{cons of mass}
    \sum_{i}c_i(x,y,t) = 1
\end{equation}
The bulk free energy per atom is defined as $f(c_A,c_B,c_C)$:
\begin{equation}
\label{bulk f}
    f(c_A,c_B,c_C) = A_1\sum_{i\neq j}c_i^2c_j^2 + A_2\prod_{i=A,B,C}c_i^2
\end{equation}
where $i,j = A,B,C$ and $A_1, A_2$ are constants. The Cahn-Hilliard free energy functional ~\cite{Cahn-Hilliard} gives the total free energy F for an isotropic, three component system as:
\begin{equation}
\label{Functional}
    F = N_v\int_V\Bigg[f(c_A,c_B,c_C) + \sum_{i=A,B,C}K_i(\nabla c_i)^2\Bigg]
\end{equation}
where $N_v$ is the number of atoms per unit volume. As the minima of $f(c_A, c_B, c_C)$ are at the extreme compositions, the bulk free energy component drives the phase separation. Jumps in concentration are simultaneously penalised by the $K_i(\nabla c_i)^2$ component in the functional $F$ to prevent sudden interfaces. As a result, a smooth or diffuse interface rather than a sharp concentration jump is obtained. The presence of the gradient energy coefficients $K_A, K_B \& K_C$ corrects these gradients. The following are the governing equations for the evolution of the material's microstructure:

\begin{equation}
\label{evol cB}
    \frac{\partial c_B}{\partial t} = M_{BB}\bigg[\nabla^2\bigg(\frac{\partial f}{\partial c_B}\bigg)-2K_{BB}\nabla^4c_B - 2K_{CB}\nabla^4c_C\bigg]
    + M_{BC}\bigg[\nabla^2\bigg(\frac{\partial f}{\partial c_C}\bigg)-2K_{CC}\nabla^4c_C - 2K_{CB}\nabla^4c_B\bigg]
\end{equation}

\begin{equation}
\label{evol cC}
    \frac{\partial c_C}{\partial t} = M_{CC}\bigg[\nabla^2\bigg(\frac{\partial f}{\partial c_C}\bigg)-2K_{CC}\nabla^4c_C - 2K_{BC}\nabla^4c_B\bigg]
    + M_{BC}\bigg[\nabla^2\bigg(\frac{\partial f}{\partial c_B}\bigg)-2K_{BB}\nabla^4c_B - 2K_{BC}\nabla^4c_C\bigg]
\end{equation}
 where, $K_{CC}= K_C+K_B;K_{BB}=K_B+K_A$ and $K_{BC}=K_{CB}=K_A$. Note that $M_{ij}$ are the mobilities, which we also consider to be independent of composition fields, and are within the laboratory frame of reference. 

The semi-implicit Fourier spectral approach, first utilised by Chen et al. \cite{CHEN1998147}, is then applied. This converts the Eqns. \ref{evol cB} and \ref{evol cC} into ordinary differential equations in the Fourier space by performing a Fourier transform on them. Applying backward difference, we obtain the following equations:


\begin{equation}
\label{cahat}
    \hat{c_C}(\textbf{k},t+\Delta t) = \frac{\hat{c}_C(\textbf{k},t) - k^2\Delta t[M_{CC}\hat{g}_C(\textbf{k},t)+M_{CB}\hat{g}_B(\textbf{k},t)]-2k^4\Delta t\hat{c}_B(\textbf{k},t)[M_{CC}K_{CB}+M_{CB}K_{BB}]}{1+2\Delta tk^4(M_{CC}K_{CC}+M_{CB}K_{CB})} 
\end{equation}

\begin{equation}
\label{cbhat}
    \hat{c_B}(\textbf{k},t+\Delta t) = \frac{\hat{c}_B(\textbf{k},t) - k^2\Delta t[M_{BB}\hat{g}_B(\textbf{k},t)+M_{CB}\hat{g}_C(\textbf{k},t)]-2k^4\Delta t\hat{c}_C(\textbf{k},t)[M_{BB}K_{CB}+M_{CB}K_{CC}]}{1+2\Delta tk^4(M_{CC}K_{BB}+M_{CB}K_{CB})} 
\end{equation}

where, \textbf{k} is the reciprocal lattice vector and $k=|\textbf{k}|$. $\hat{c}_C(\textbf{k},t)$ and $\hat{c}_B(\textbf{k},t)$ are the Fourier transforms of the respective compositions in real space. We define the Fourier transforms of the bulk driving force  $g_C,g_B$ as $\hat{g}_C$, $\hat{g}_B$ respectively, where $g_C=(\partial f/\partial c_C)$ and $g_B=(\partial f/\partial c_B)$. In the semi-implicit formulation, we treat $\hat{c}_C(\textbf{k},t)$ and  $\hat{c}_B(\textbf{k},t)$ implicitly, while $\hat{g}_C$ and $\hat{g}_B$ are treated explicitly. Eqns.~\ref{cahat} and \ref{cbhat} are solved iteratively to get the temporal evolution of the microstructure.

\subsection{Dataset}
Since there exist two independent evolving fields in the problem, $c_B$, $c_C$, the neural network based solver must be capable of simultaneously learning the dynamics of the evolution of both the concentration fields and the relation between the two. Hence, an effective field variable is defined, $u_{T:T+N}$, that takes into account the time-series of both $c_B$ and $c_C$ for N time-steps from T to T+N. The field variable $u_{T:T+N}$ is defined by stacking the concentration maps of B and C after discretization on a mesh of size $d_x$ by $d_y$ resolution. The shape of $u_T$ that is fed into the model then becomes \{N, $d_x$, $d_y$, 2\}. Based on the frequency of saving outputs from the numerical solver, the time difference between two consecutive time-steps in the data relative to the numerical solver is denoted by $d\tau$. For the purpose of this study, 1000 possible evolution trajectories are simulated using the numerical solver for a square mesh of dimensions $d_x = d_y = 64$ for 5 different values of $d\tau$ which are 1, 20, 40, 200, and 500, each. These values of $d\tau$ were selected to understand the model's performance during different stages of the microstructure evolution, using $d\tau=1$ for the early stages of phase seperation, $d\tau=20, 40, 200$ for the intermediate stages and $d\tau=500$ for the late stages of grain coarsening. 

\begin{figure}
     \centering
     \begin{subfigure}[b]{\textwidth}
         \centering
         \includegraphics[width=0.4\textwidth]{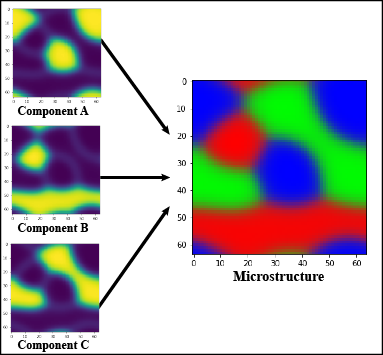}
         \caption{Microstructure map formed from the individual concentration fields.}
     \end{subfigure}
     \hfill
     \begin{subfigure}[b]{\textwidth}
         \centering
         \includegraphics[width=0.4\textwidth]{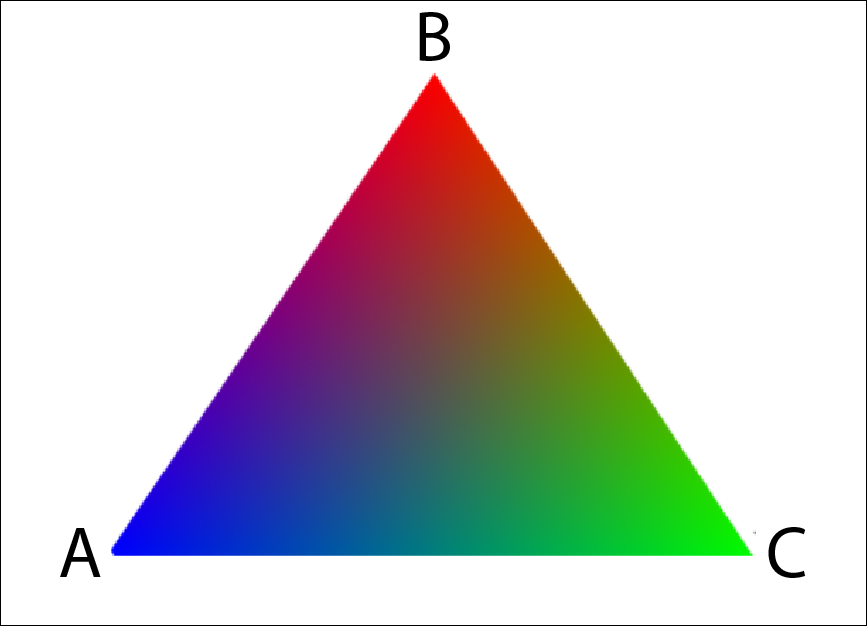}
         \caption{RGB map projected onto the Gibbs triangle.}
         \label{fig:cmap}
     \end{subfigure}
     \hfill
        \caption{There exist three concentration fields in ternary systems, one for each of the components A, B and C of which only two are independent. At any instant, the three concentration fields can be stacked on top of each other, similar to how the three channels of an RGB image are stacked, to visualize the full microstructure. As can be inferred from Figure \ref{fig:cmap}, here the A-rich phase has been shown in Blue, B-rich phase in Red and C-rich phase in Green.}
	\label{fig:repfig}
\end{figure}

\subsection{Generative Adversarial Networks}
\begin{figure}
	\centering 
	\includegraphics[width=\textwidth]{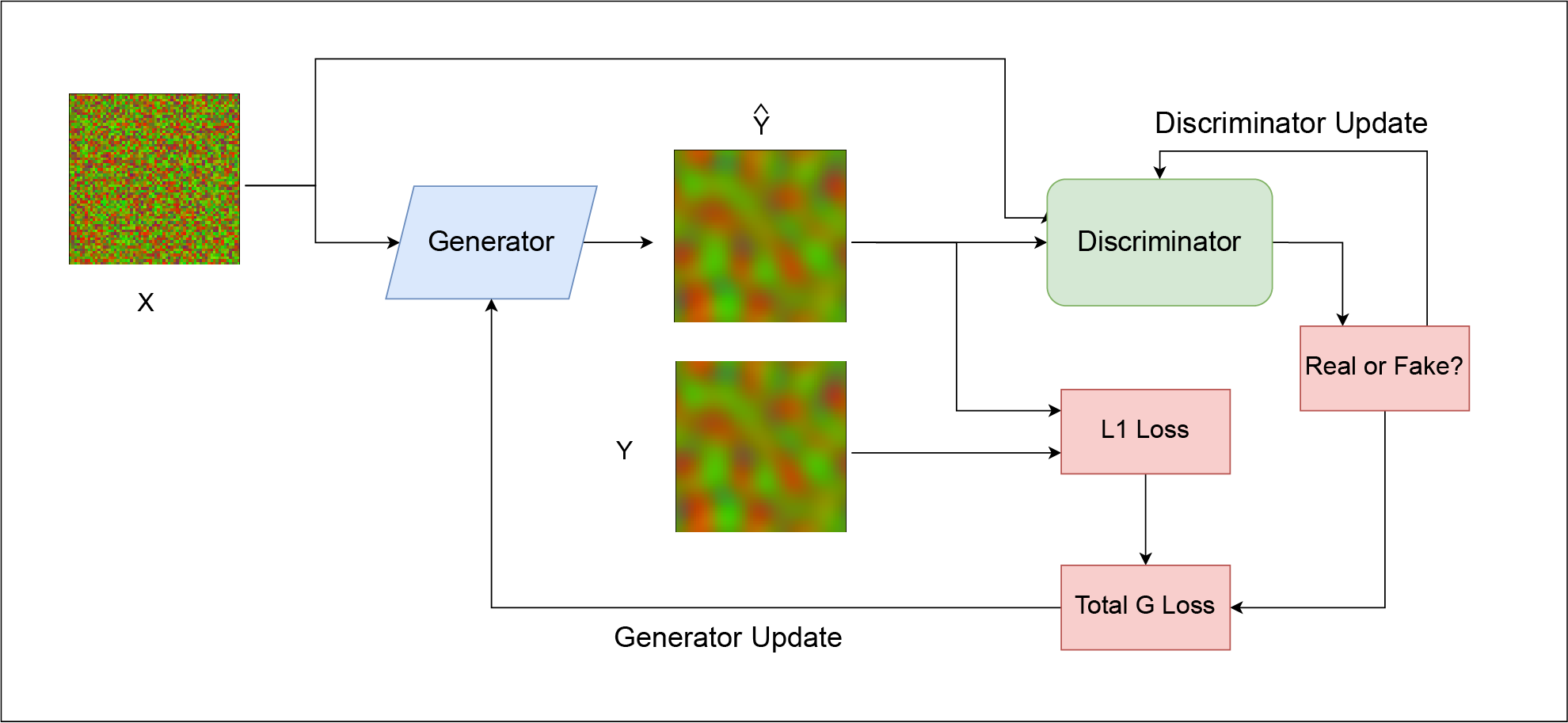}
	\caption{A cGAN model uses initial concentration fields as input, which are denoted by $X$, to the generator, which then attempts to generate future concentration fields at prescribed time intervals (output of the generator), which are denoted by $\hat{Y}$ and the ground truth of the fields are denoted by $Y$ . The discriminator takes both the input and output images and detects whether the output field is ``Real'' or ``Fake''. The loss associated with the discriminator is obtained from the misclassification of images into Real or Fake categories, while the loss associated with the generator is a weighted sum of the $L_1$ loss and the loss of the discriminator. }
	\label{fig:framework}
\end{figure}

The Generative Adversarial Network (GAN) \cite{goodfellow2014generative} has shown great success since its conception towards learning high-level features from a given distribution and generating new samples that are indistinguishable from the real data in an unsupervised manner. A GAN consists of two deep Convolutional Neural Networks, called the Generator and the Discriminator, that are trained in an adversarial manner. The generator learns the mapping between an input random noise and the given real data distribution to generate the fake samples, while the discriminator learns to classify samples as being real or fake. Here, real refers to data that is taken from the actual data distribution while fake refers to the data created by the Generator. Hence, $G: Z \rightarrow W$ and $D: W \rightarrow [0, 1]$, where $Z$ is a Gaussian prior distribution and $W$ is the distribution of real data. Both $D$ and $G$ are trained simultaneously, and the learning happens in a min-max fashion, with the discriminator attempting to minimize its classification loss and the generator attempting to fool the discriminator by maximizing the discriminator's loss. The training objective is given by \cite{goodfellow2014generative}: 
\begin{equation}
\label{Eqn1}
    \min _{G} \max _{D} \mathbb{E}_{\boldsymbol{w} \sim p_{\text {data}}(\boldsymbol{w})}[\log D(\boldsymbol{w})]+\mathbb{E}_{\boldsymbol{z} \sim p_{\boldsymbol{w}}(\boldsymbol{z})}[\log (1-D(G(\boldsymbol{z})))]
\end{equation}
A modification to the GAN architecture, called Pix2Pix, involving the use of an encoder-decoder based U-net generator and a patch discriminator was proposed to allow for effective Image-to-Image translation \cite{isola2018imagetoimage}. In this case, the generator takes an image from one domain as the input ($X$) and learns the mapping to an image from the target domain ($Y$), i.e., $G: X \rightarrow Y$. The objective function for such a GAN is given in Equation \ref{Eqn2}. The Pix2Pix GAN framework is shown pictorially in Figure \ref{fig:framework}.
\begin{equation}
\label{Eqn2}
    \mathcal{L}_{c G A N}(G, D)= \mathbb{E}_{x, y}[\log D(x, y)]+\mathbb{E}_{x}[\log (1-D(x, G(x))]
\end{equation}
\begin{figure}
	\centering
	\includegraphics[width=0.8\textwidth]{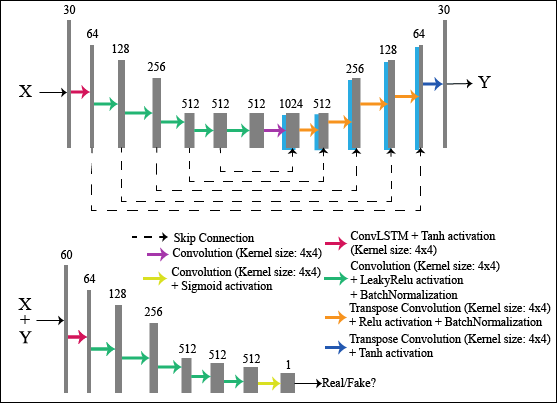}
	\caption{Schematics of the generator (Top) and discriminator (Bottom) for the proposed conv-LSTM GAN model. The number of feature maps are indicated above the output of each block. The distinctive ``U"-shaped encoder-decoder architecture for the generator can be clearly seen in the schematic. The input layer in the discriminator has twice the number of filters used in the encoding part of the generator. The blue blocks in the decoding part of the network are the outputs of the corresponding encoding block obtained via the skip connections which are represented as dashed lines, which are concatenated before the corresponding decoding block. The U-net GAN model has approximately the same architecture as shown in this schematic, with convolutional layers in place of the conv-LSTM layers.}
	\label{fig:archfig}
\end{figure}
\subsection{Model Description}
We cast the problem of predicting the microstructure evolution as an Image to Image translation problem. We propose two possible approaches to learn the operator that describes the coupled PDEs that were mentioned in Section \ref{Problem Formulation}. The first uses a fully convolutional U-net GAN to learn the operator that maps $u_{t}$ to $u_{t+1}$, and the second uses a modified U-Net GAN with a conv-LSTM layer to learn the map from $u_{t:t+N}$ to $u_{t+N:t+M+N}$, where $N, M \geq 1$. To allow for ease of identification, the first model is referred to as simply "U-net GAN" and the second as "conv-LSTM GAN". The former was chosen to demonstrate the ability of such GANs to learn the operator mapping corresponding to coupled PDEs. The latter builds on this by taking the temporal nature of the data into account in a recurrent manner and therefore allows to predict longer sequences of the evolution process in one go. The only difference between the two proposed models' architecture is the first layer of both the generator and discriminator network. In the U-net GAN, this is a convolution layer with a LeakyRelu activation while in the other model, a conv-LSTM layer with tanh activation is used instead. 

The generator and discriminator’s architecture for the conv-LSTM GAN is shown in Fig \ref{fig:archfig}. In this case, the generator consists of a conv-LSTM layer with tanh activation followed by 4 encoding blocks and 4 decoding blocks with a convolution layer at the bottleneck. Skip connections are provided from each encoding block to its corresponding decoding block making the network resemble a U shape. Since the change in the overall structure of the microstructure between consequent time-steps is expected to minimal, the skip connections allow for the flow of low-level features between corresponding encoding and decoding blocks. Each encoding block consists of a convolution layer having a stride of 2 with LeakyReLU activation followed by a batch normalization layer. The decoding block is an approximate mirror of the encoder, having a transpose convolution layer with ReLU activation followed by batch normalization and dropout layers. 

The discriminator has approximately the same architecture as the encoding part of the generator, the only differences being the input images are concatenated before passing through the encoding blocks, and the last convolution block applies a sigmoid activation to produce the probability of the input pair to be from the actual distribution of training data. All convolution and transposed convolution operations use a kernel of size 4 by 4 and have padding applied as well.

In contrast to similar works in the past, since the proposed models are fully convolutional in nature and lack any fully connected layers, they are invariant to the size of the input data, i.e, independent of the size and geometry of the mesh taken while discretizing the domain. Similar to the Pix2Pix framework \cite{isola2018imagetoimage}, to ensure the data generated by the generator is accurate in the $L_1$ sense, we add a weighted $L_1$ norm, shown in Equation \ref{Eqn3} to the loss function of the GAN. The training objective for the proposed model is shown in Equation \ref{Eqn4}.
\begin{equation}\label{Eqn3}
\mathcal{L}_{L_1}(G)=\mathbb{E}_{x, y}\left[\|y-G(x)\|_{1}\right]
\end{equation}
\begin{equation}\label{Eqn4}
G^{*}=\arg \min _{G} \max _{D} \mathcal{L}_{c G A N}(G, D)+\lambda \mathcal{L}_{L_1}(G)
\end{equation}
Here, $x \in X$, $y \in Y$ and $\lambda$ is a hyperparameter that weighs in the $L_1$ component of the loss function. For simplicity, the GAN component of the loss can be thought of as a learned loss function that quantifies the realness of the data generated while training, which takes into account the overall features present in the outputs of the generator. Meanwhile, the $L_1$ component of the loss ensures pixel-wise accuracy. 

\subsection{Training Configurations}
\begin{figure}
	\centering
	\includegraphics[width=0.6\textwidth]{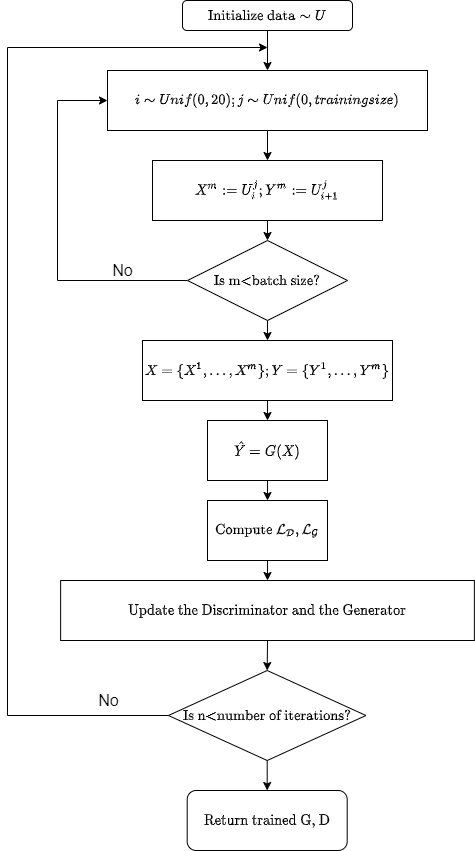}
	\caption{cGAN training algorithm using mini-batch Stochastic Gradient Descent, adapted from \cite{goodfellow2014generative}. We set the mini-batch size, $m$, to be 1. In practice, for the U-net GAN sampling the data for a random transition, $u_{i:i+1}$, is found to work better than using only a specific transition, say $u_{0:1}$, as input. This step is omitted for the conv-LSTM GAN and a series of $u_{T:T+15}$ is directly fed to the model to utilize the temporal nature of the data as well. Furthermore, $\hat{Y}$ is the prediction made by the model, which is used to compute the losses during training.}
	\label{alg1}
\end{figure}

\begin{figure}
     \centering
     \begin{subfigure}[b]{6.5cm}
         \centering
         \includegraphics[width=\textwidth]{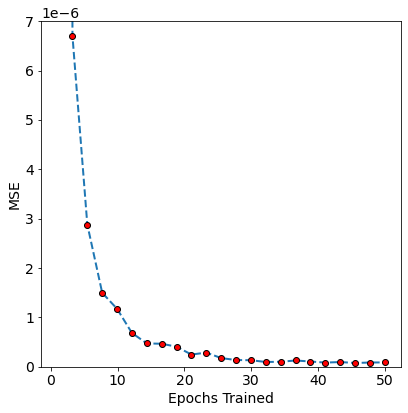}
         \caption{Variation of Mean Squared Error (MSE) with number of epochs used for training, line is drawn as a guide to the eye.}
         \label{fig:train_loss}
     \end{subfigure}
     \hfill
     \begin{subfigure}[b]{6.75cm}
         \centering
         \includegraphics[width=\textwidth]{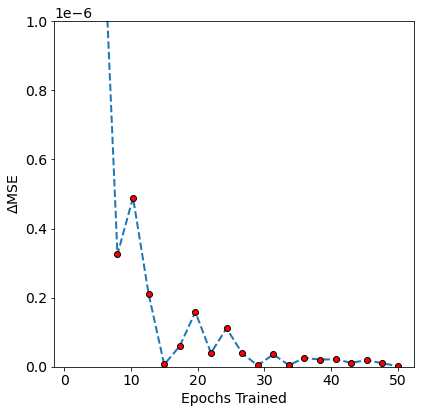}
         \caption{Variation of change in Mean Squared Error (MSE) with number of epochs used for training, line is drawn as a guide to the eye.}
         \label{fig:del_train_loss}
     \end{subfigure}
     \hfill
        \caption{As can be seen, the MSE starts to flatten out quite early in the training (within 20 epochs), and the difference in the MSE of subsequent epochs falls to below $10^{-7}$ after 30 epochs, and below $10^{-8}$ after 45 epochs of training. Hence, setting the maximum number of epochs of training to 50 can be expected to suffice while training the proposed models on our datasets.}
        \label{fig:metrics_evol}
\end{figure}

Both models were trained using the algorithm shown in Figure \ref{alg1}, which was adapted from the algorithm suggested by \cite{goodfellow2014generative}. The U-net GAN was trained on data with $d\tau=200$ (referred to as $\mathcal{M}_0$), while the conv-LSTM GAN was trained on 4 different datasets, having $d\tau=1, 20, 40, 500$ (referred to as $\mathcal{G}_1$, $\mathcal{G}_{20}$, $\mathcal{G}_{40}$, $\mathcal{G}_{500}$ respectively). Each data point was normalized to the range of -1 to 1 to allow for stable training of the GANs \cite{salimans2016improved}. To reiterate, the models learn the mapping between $u_{t:t+N}$ and $u_{t+N:t+N+M}$, i.e., the consolidated concentration maps of components B and C of the material for N time-steps between time t and t+N are used as input and the model is made to predict the next M time-steps. For the purpose of this study, the two models proposed use N equal to 0 and M equal to 1 for the U-net GAN and M equal to 15 for the conv-LSTM GAN. Therefore, the domains $X$ and $Y$ being mapped by the network as per Equation \ref{Eqn2} are defined as:
\begin{equation}
    u_{t:t+15}^i = [u_t^i, u_{t+1}^i, ..., u_{t+14}^i] \in X
\end{equation}
\begin{equation}
    u_{t+15:t+30}^i = [u_{t+15}^i, u_{t+16}^i, ..., u_{t+29}^i] \in Y
\end{equation}
Here, $u_t^i$ = $[c_B^t, c_C^t]^i$, which is the $i^{th}$ training sample of the dataset and $\mathcal{G}: X \rightarrow Y$, where $\mathcal{G}$ is the model used. The $\lambda$ hyperparameter in the loss function was set as 100, after optimizing in the range of $(50, 150)$. The learning rate of the model was taken as 0.0002 and decayed to 0.00001 during training and the Adam optimizer \cite{kingma2017adam} was used. A total of 900 training samples were provided to the model while training in batches of size 1 and 100 samples were left out for testing. It was found that increasing the batch size led to a reduction in the quality of the generated images with regards to the morphological features and the resolution.  The model was implemented in the TensorFlow Deep-Learning framework \cite{tensorflow2015-whitepaper} using the Keras API \cite{chollet2015keras}. The variation of the mean squared error (MSE) with the number of epochs of training of the conv-LSTM GAN model is shown in Figure \ref{fig:train_loss}. Based on this variation and visual inspection during training, the length of training was fixed to 50 epochs for the models, since there was very little improvement in the MSE, as can be seen from Figure \ref{fig:del_train_loss} and this is expected to be sufficient for the proposed models to converge. The models were trained for 50 epochs on both, a single Tesla K80 GPU with 12GB Memory and a Tesla V100 GPU, and the training times for the two GPUs are shown in Table \ref{tab:table_times}.
\begin{table}
	\caption{Training times (in seconds) per epoch for both the U-net GAN ($\mathcal{M}_0$) and the conv-LSTM GAN models ($\mathcal{G}$) on the available GPUs.}
	\centering
	\begin{tabular}{lcc}
    \hline \multicolumn{1}{c}{Model} & K80 & V100 \\
    \hline $\mathcal{M}_0$ & $965.5708$ & $714.5753$ \\
    $\mathcal{G}$ & $705.2787$ & $498.5840$ \\
    \hline
    \end{tabular}
	\label{tab:table_times}
\end{table}

\section{Results and Discussion}
In this section, we first understand the ability of the U-net GAN model to learn the operator corresponding to the coupled Cahn-Hilliard equations. Following this, we focus our analysis on the conv-LSTM GAN model, owing to its ability to take the temporal nature of the data into account. Any Neural-PDE solver must be accurate and generalizable to new inital configurations, must be computationally feasible, offer a speed-up over existing numerical methods and should also preferably be scalable and transferable to a new mesh without retraining. To quantify the errors made by the models, the mean squared error (MSE) and structural similarity index (SSIM) were used. Hence, we analyze these aspects of operator training using the conv-LSTM GAN model and also comment on the effect of changing the value of $d\tau$ of the data that is fed to the model. 

\subsection{Learning the operator using U-net GANs}
\begin{figure}
     \centering
     \begin{subfigure}[b]{\textwidth}
         \centering
         \includegraphics[width=0.5\textwidth]{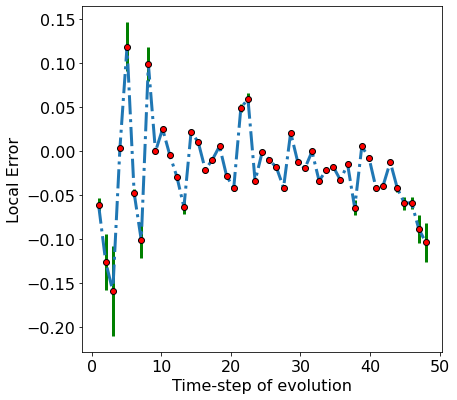}
         \caption{Local error when predicting between subsequent time-steps, which is defined as per Equation 7. The vertical green bars represent the standard deviation of the error across the test set.}
         \label{fig:localerr}
     \end{subfigure}
     \hfill
     \begin{subfigure}[b]{\textwidth}
         \centering
         \includegraphics[width=0.9\textwidth]{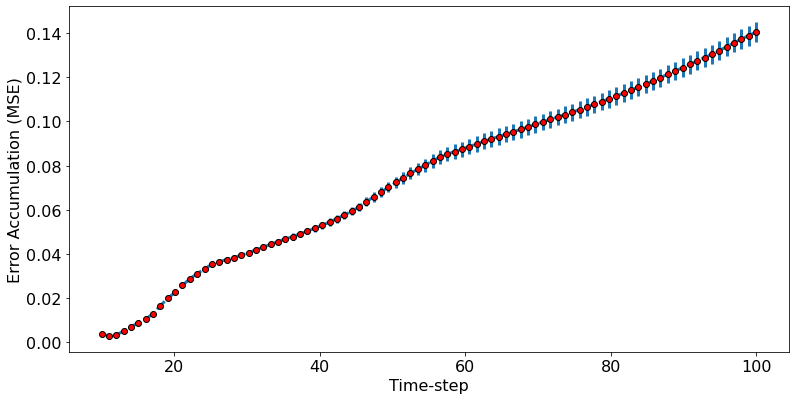}
         \caption{
         The change in error accumulation (measured using the mean squared error) when recursively predicting using the model $\mathcal{M}_0$. The vertical blue bars represent the standard deviation of the MSE across the test set.}
         \label{fig:globalerr}
     \end{subfigure}
     \hfill
        \caption{The U-net GAN ($\mathcal{M}_0$) is capable of making predictions for a single time-step. Hence, here the predictions were made for $u_{t+1}$ given $u_t$, and the variation of the local and global errors (along $y$ in both plots) with $t$ (along $x$ in both plots) has been shown. The MSE and SSIM of the predictions made by the U-net GAN are quite poor when predicting the early time-steps of evolution, and both metrics significantly improve after the $8^{th}$ time-step.}
        \label{fig:metrics_1-1}
\end{figure}

\begin{figure}
	\centering
	\includegraphics[width=0.75\textwidth]{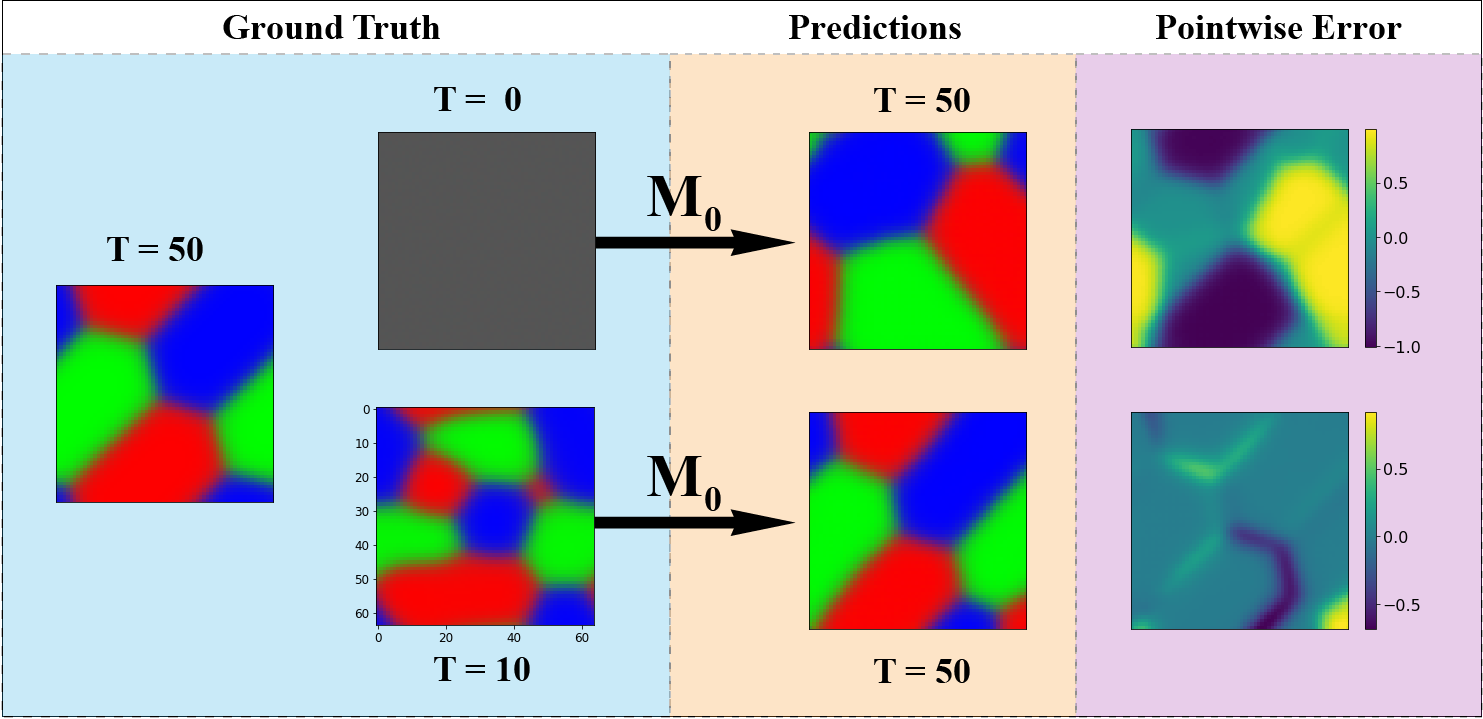}
	\caption{The trained U-net GAN model ($\mathcal{M}_0$) was made to recursively predict the near-equilibrium microstructure (at T=50) starting from different steps of the evolution, with $d\tau = 200$. The predictions made by the model are far closer to the ground truth at T=50 when the predictions were started from later time-steps ($T\geq10$). Although the pixel-wise difference in the predictions of the near-equilibrium solution is considerable when starting from an early time-step, the microstructure predicted does appear plausible, showing the correct morphology of the grains. Blue Region: Ground Truth, Red Region: Predictions by the model.}
	\label{fig:res_op}
\end{figure}

\begin{figure}
    \centering
    \includegraphics[width=\textwidth]{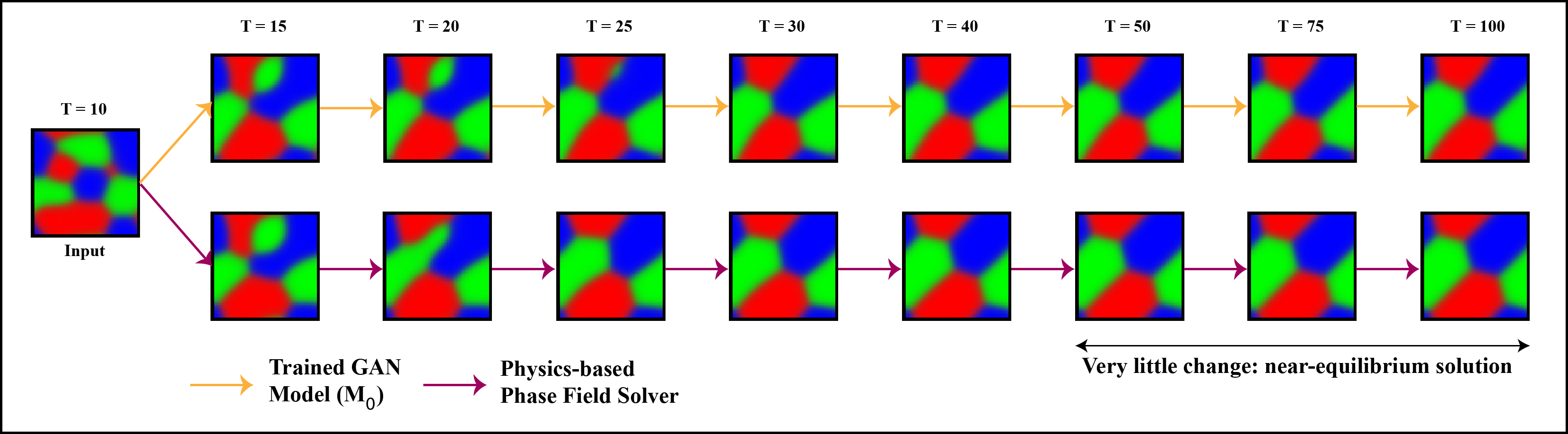}
    \caption{Evolution trajectories obtained when recursively predicting, starting from the $10^{th}$ step of evolution ($d\tau = 200$), using the trained model $\mathcal{M}_0$ (Top), and the physics-based solver (Bottom).}
    \label{fig:recursive}
\end{figure}
Since the U-net GAN ($\mathcal{M}_0$) has been trained to learn the operator corresponding to a single time-step, in order to predict the complete evolution trajectory we recursively make predictions. One such evolution trajectory has been shown in Figure \ref{fig:recursive}. To understand the accumulation of error in the solution when recursively predicting the evolution trajectory, we use define the local error and global truncation errors, and their variation with the time step of evolution predicted have been shown in Figure \ref{fig:metrics_1-1}.
\begin{equation}
    \text{Local Error} = \frac{MSE_{t+\delta t} - MSE_{t}}{MSE_{t}}
\end{equation}
Here, $MSE_t$ is the mean squared error of the prediction with the ground truth at time-step $t$. The variation of the local error provides an insight into how well the model can predict the evolution of the system from time-step $t$ to $t+\delta t$. As can be seen from Figure \ref{fig:localerr}, there is a considerable reduction in the local error in the earlier time-steps (0 to 10) in comparison to the later ones (10 to 50), implying the model being better at predicting the evolution trajectory during later time-steps. The variation of global error when recursively predicting from the $10^{th}$ time-step, shown in Figure \ref{fig:globalerr}, appears to be flatten out with increasing time, showing that the error accumulation is largely in the early steps of evolution. Furthermore, the model was made to recursively predict the near-equilibrium microstructure of the system from different starting points and the quality of the resulting predictions can be visually seen in Figure \ref{fig:res_op}. The metrics (MSE and SSIM) for the error in the final solution have been tabulated in Table \ref{tab:table_1-1}. As can be seen, the final solution predicted by the trained model starting from the initial conditions ($T=0$) is very poor with regards to the metrics used, and the error metrics for the predictions improve when starting from a later time-step, which is to be expected. However, on closer inspection of Figure \ref{fig:res_op}, the morphology (shape and sizes of the grains) of the microstructure predicted starting from $T=0$ is visually very similar to the final solution, albeit the predicted grain orientations are incorrect. If one was to not know the actual solution, the predictions made by the model appear plausible. This can be attributed to the chaotic dynamics of the evolution in the first 3-4 time-steps, wherein the grains corresponding to distinct phases form. Hence, even small errors in prediction in this chaotic regime blow up considerably with time and lead to a very different near-equilibrium solution. We hypothesise that the trained model is capable of predicting the near-equilibrium solution to the problem when starting after this particular regime, i.e., after distinct grains have formed in the microstructure, which happens around the $3^{rd}-4^{th}$ step ($600-800^{th}$ step for the numerical solver) of the evolution. The plots in Figure \ref{fig:metrics_1-1} and the predictions made after $T=5$ which can be seen in Figure \ref{fig:res_op} support this hypothesis. These predictions made by the model can in part be attributed to the generation capabilities of GANs using their ability to learn the distribution of the data and generate new ``fake'' samples that lie within this distribution. Therefore, we believe the U-net GAN model has effectively learnt the operator that corresponds to one time-step of evolution after distinct grains have been formed in the microstructure. Knowing this, we then extend this approach of using a U-net GAN to learn the operator that describes the evolution of longer sequences. This is precisely the goal of the conv-LSTM GAN ($\mathcal{G}_1, \mathcal{G}_{20}, \mathcal{G}_{40} \& \mathcal{G}_{500}$), and the next sections are devoted towards analyzing various aspects of the predictions made by this recurrent model.

\begin{table}
	\caption{Results of predictions for the near-equilibrium solution made by the trained U-net GAN model starting from the $T^{th}$ step of the evolution}
	\centering
	\begin{tabular}{lcccccc}
    \hline \multicolumn{1}{c}{Metric} & T=0 & T=5 & T=10 & T=20 & T=30 & T=40\\
    \hline MSE & $4.841\times10^{-1}$ & $4.093\times10^{-2}$ & $2.41\times10^{-2}$ & $7.2\times10^{-3}$ & $8.7\times10^{-3}$ & $2.3\times10^{-3}$\\
    SSIM & $0.00525$ & $0.66324$ & $0.7442$ & $0.9274$ & $0.9293$ & $0.9657$\\
    \hline
    \end{tabular}
	\label{tab:table_1-1}
\end{table}
\subsection{Accuracy and Generalizability}
The trained conv-LSTM GAN model was made to predict on the data left out for testing and the predictions made have been compared to the ground truth in Figure \ref{fig:Fig4}. The results of these metrics have been consolidated in Table \ref{tab:table1}. To understand this model’s extrapolation ability, predictions were recursively made for the next 15 time steps and the results have also been shown in Fig \ref{fig:Fig4}. To assess the model’s interpolation ability, predictions were made by passing data from time-steps 5 to 20 to the model and the results are shown in Fig \ref{fig:Fig5}. As can be inferred from Table \ref{tab:table1} and Figure \ref{fig:Fig4}, this model is capable of predicting the evolution very accurately when the time difference between subsequent inputs to the model, i.e, $d\tau$ is 1. Further, as is evident from Fig \ref{fig:Fig5}, the model is also able to interpolate within the given time series, i.e, it can start making predictions from an intermediate time-step as well. Although the model has seen only the first 30 time-steps during training, it is capable of extrapolating ahead fairly accurately as well.
\begin{figure}
     \centering
     \begin{subfigure}[b]{6cm}
         \centering
         \includegraphics[width=\textwidth]{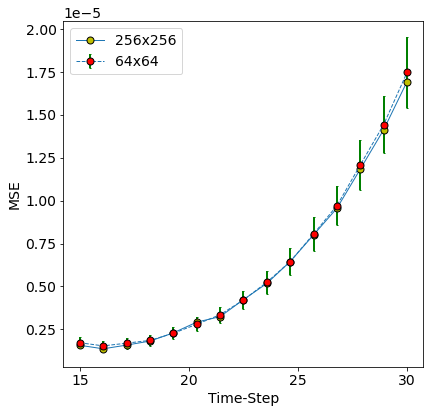}
         \caption{Plot between MSE and time-step for $\mathcal{G}_1$}
         \label{fig:MSE_T1}
     \end{subfigure}
     \hfill
     \begin{subfigure}[b]{6cm}
         \centering
         \includegraphics[width=\textwidth]{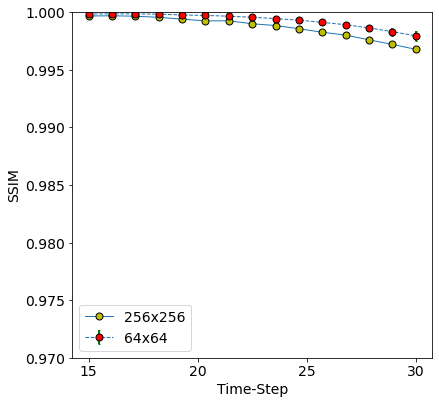}
         \caption{Plot between SSIM and time-step for $\mathcal{G}_1$}
         \label{fig:SSIM_T1}
     \end{subfigure}
     \hfill
     \begin{subfigure}[b]{6cm}
         \centering
         \includegraphics[width=\textwidth]{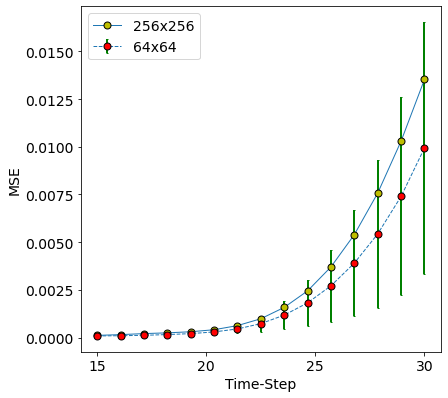}
         \caption{Plot between MSE and time-step for for $\mathcal{G}_{20}$}
         \label{fig:MSE_T20}
     \end{subfigure}
     \hfill
     \begin{subfigure}[b]{6cm}
         \centering
         \includegraphics[width=\textwidth]{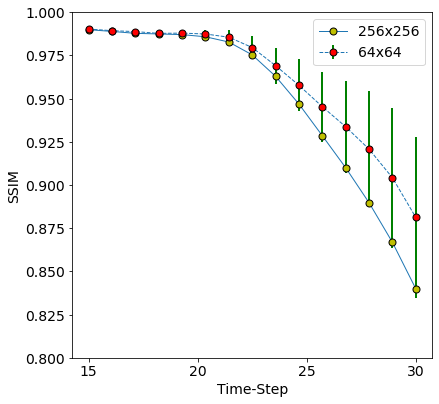}
         \caption{Plot between SSIM and time-step for $\mathcal{G}_{20}$}
         \label{fig:SSIM_T20}
     \end{subfigure}
     \hfill
     \begin{subfigure}[b]{6cm}
         \centering
         \includegraphics[width=\textwidth]{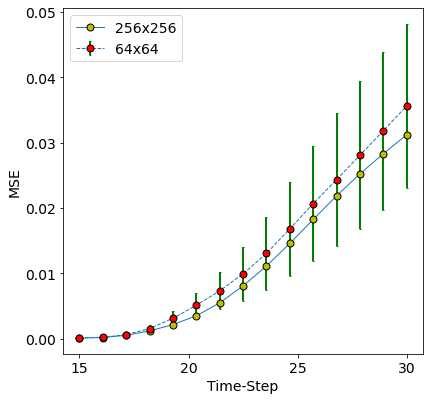}
         \caption{Plot between MSE and time-step for for $\mathcal{G}_{40}$}
         \label{fig:MSE_T40}
     \end{subfigure}
     \hfill
     \begin{subfigure}[b]{6cm}
         \centering
         \includegraphics[width=\textwidth]{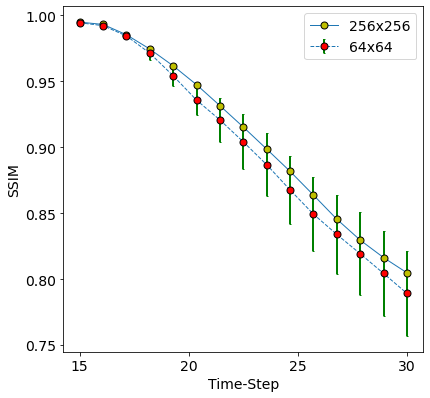}
         \caption{Plot between SSIM and time-step for $\mathcal{G}_{40}$}
         \label{fig:SSIM_T40}
     \end{subfigure}
     \hfill
        \caption{The plots of MSE and SSIM with the time-step of the predictions by the conv-LSTM GAN on both 64x64 and 256x256 mesh for $\mathcal{G}_1$ in a, b, $\mathcal{G}_{20}$ in c, d and $\mathcal{G}_{40}$ in e, f. The vertical Green bars represent the standard deviation of the MSE and SSIM on the test dataset. As can be seen, the metrics on the 256x256 mesh is approximately the same as that on the 64x64 mesh.}
        \label{fig:T1_metrics}
\end{figure}
\begin{table}
	\caption{Results of predictions made by the trained conv-LSTM GAN model for $d\tau$ values of 1, 20 and 40.}
	\centering
	\begin{tabular}{lcccccc}
    \hline \multicolumn{1}{c}{ Model Configuration } & Metric & T=15 & T=29 & T=30 & T=44 \\
    \hline $d\tau=1, d_x=d_y=64$ & MSE & $1.83\times10^{-6}$ & $1.53\times10^{-5}$ & $5.09\times10^{-4}$ & $1.18\times10^{-3}$ \\
    & SSIM & $0.99978$ & $0.99861$ & $0.97589$ & $0.96924$ \\
    \hline $d\tau=20, d_x=d_y=64$ & MSE & $8.87\times10^{-5}$ & $9.9\times10^{-3}$ & $1.3\times10^{-2}$ & $9.27\times10^{-2}$ \\
    & SSIM & $0.99025$ & $0.88121$ & $0.86028$ & $0.63570$ \\
    \hline $d\tau=40, d_x=d_y=64$ & MSE & $1.22\times10^{-4}$ & $2.83\times10^{-2}$ & $2.91\times10^{-2}$ & $1.07\times10^{-1}$ \\
    & SSIM & $0.99555$ & $0.78638$ & $0.78650$ & $0.64875$ \\
    \hline $d\tau=500, d_x=d_y=64$ & MSE & $1.24\times10^{-4}$ & $1.28\times10^{-3}$ & $1.76\times10^{-3}$ & $3.64\times10^{-3}$ \\
    & SSIM & $0.99870$ & $0.98956$ & $0.98697$ & $0.97367$ \\
    \hline
    \end{tabular}
	\label{tab:table1}
\end{table}
\subsection{Scalability and Transferability}
\begin{figure}
	\centering
	\includegraphics[width=14cm]{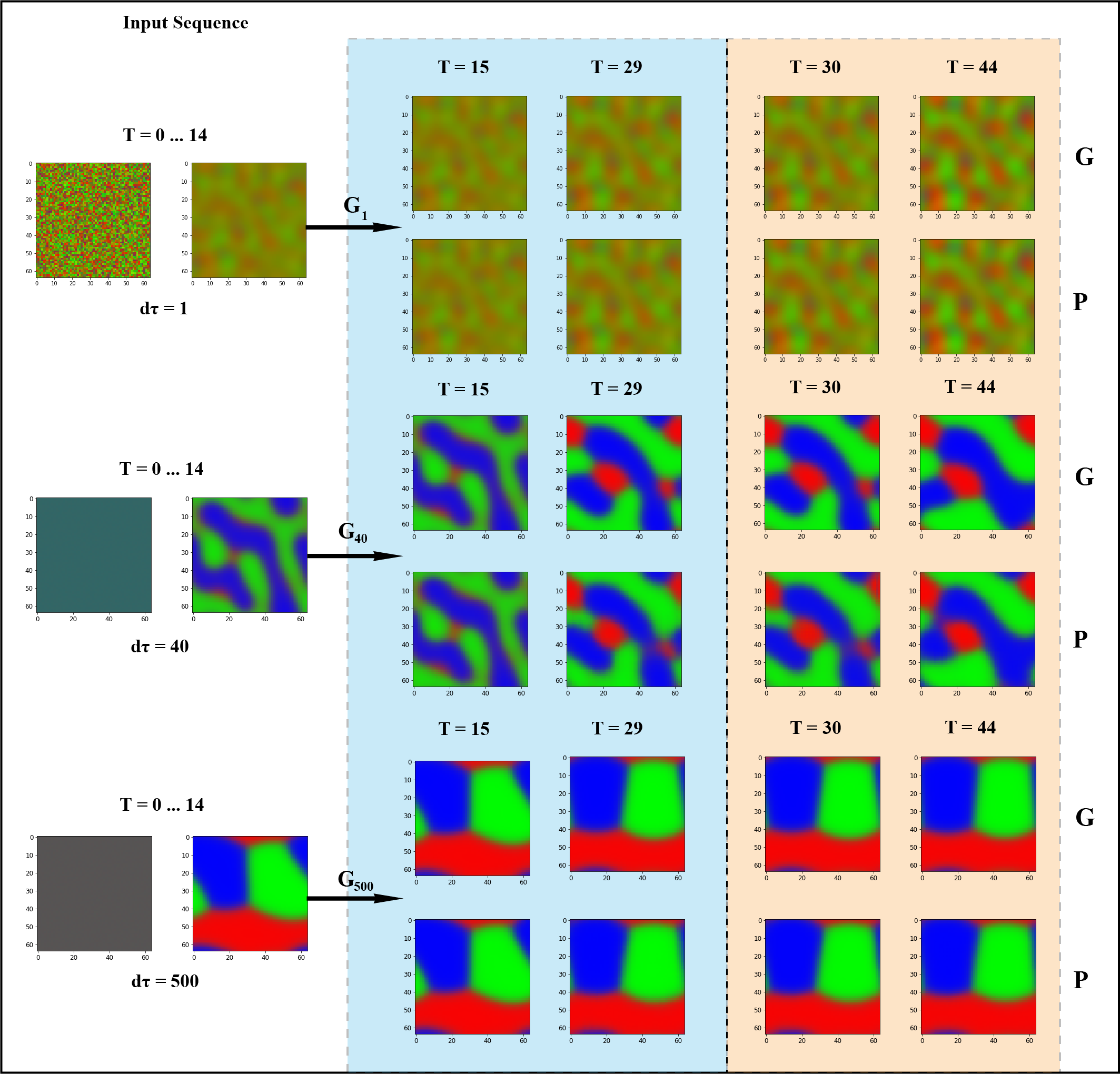}
	\caption{The sequences for the first 15 time-steps of the evolution given as input to the three of the models trained: $\mathcal{G}_1$, $\mathcal{G}_{40}$ and $\mathcal{G}_{500}$. (Left) on a domain size of 64 by 64. The Blue region (Centre) corresponds to stages of evolution the model has seen via the training data and the Red region (Right) corresponds to stages of evolution the model hasn't seen (Extrapolation regime). G: Ground Truth, P: Predictions by the models.}
	\label{fig:Fig4}
\end{figure}

\begin{figure}
	\centering
	\includegraphics[width=14cm]{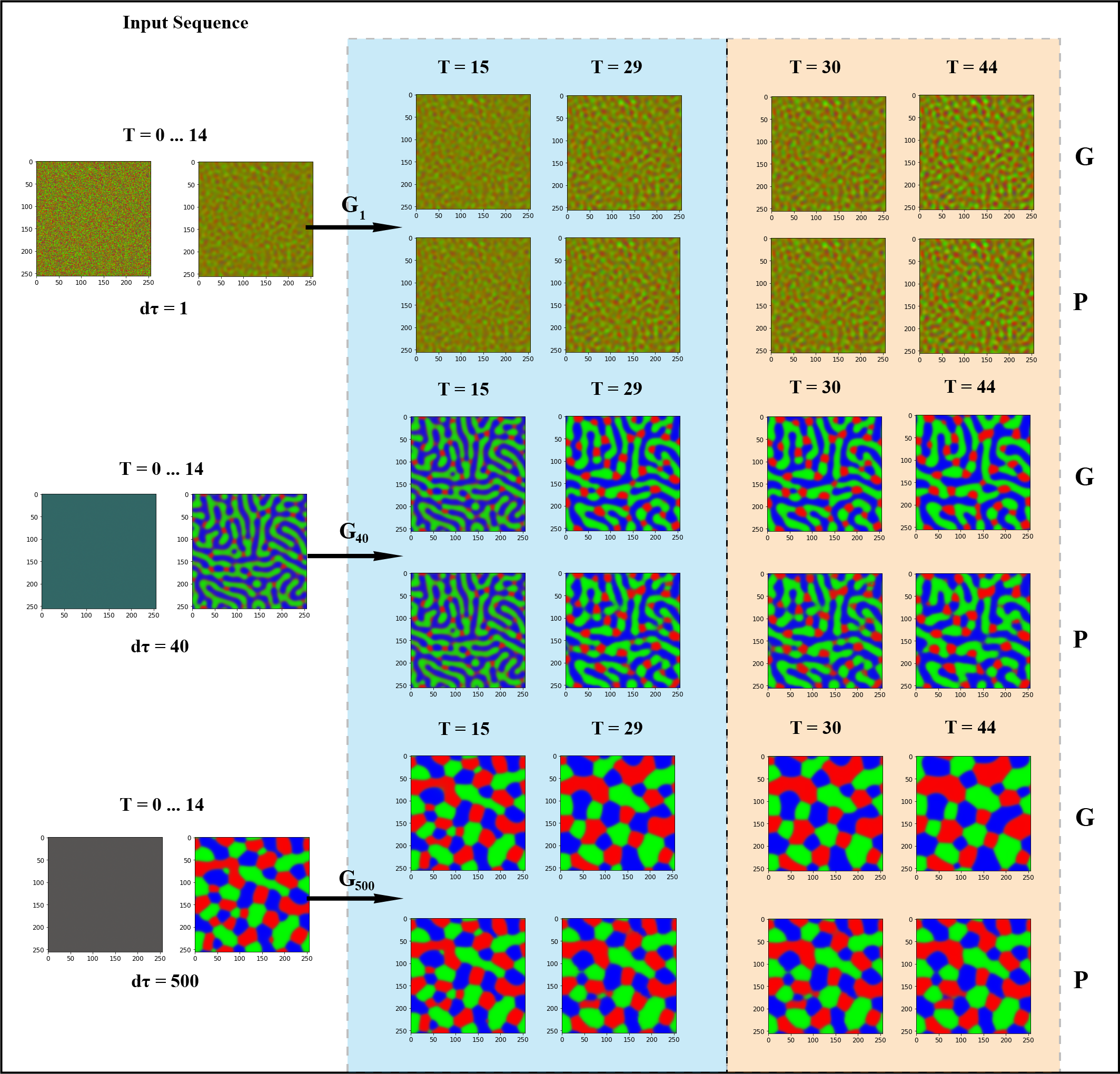}
	\caption{The sequences for the first 15 time-steps for a larger domain of size 256 by 256 are given as input to the three of the models trained: $\mathcal{G}_1$, $\mathcal{G}_{40}$ and $\mathcal{G}_{500}$. (Left) on a domain of size 64 by 64. The Blue region (Centre) corresponds to stages of evolution the model has seen via the training data and the Red region (Right) corresponds to stages of evolution the model hasn't seen (Extrapolation regime). G: Ground Truth, P: Predictions by the models.}
	\label{fig:Fig6}
\end{figure}

\begin{figure}
	\centering
	\includegraphics[width=14cm]{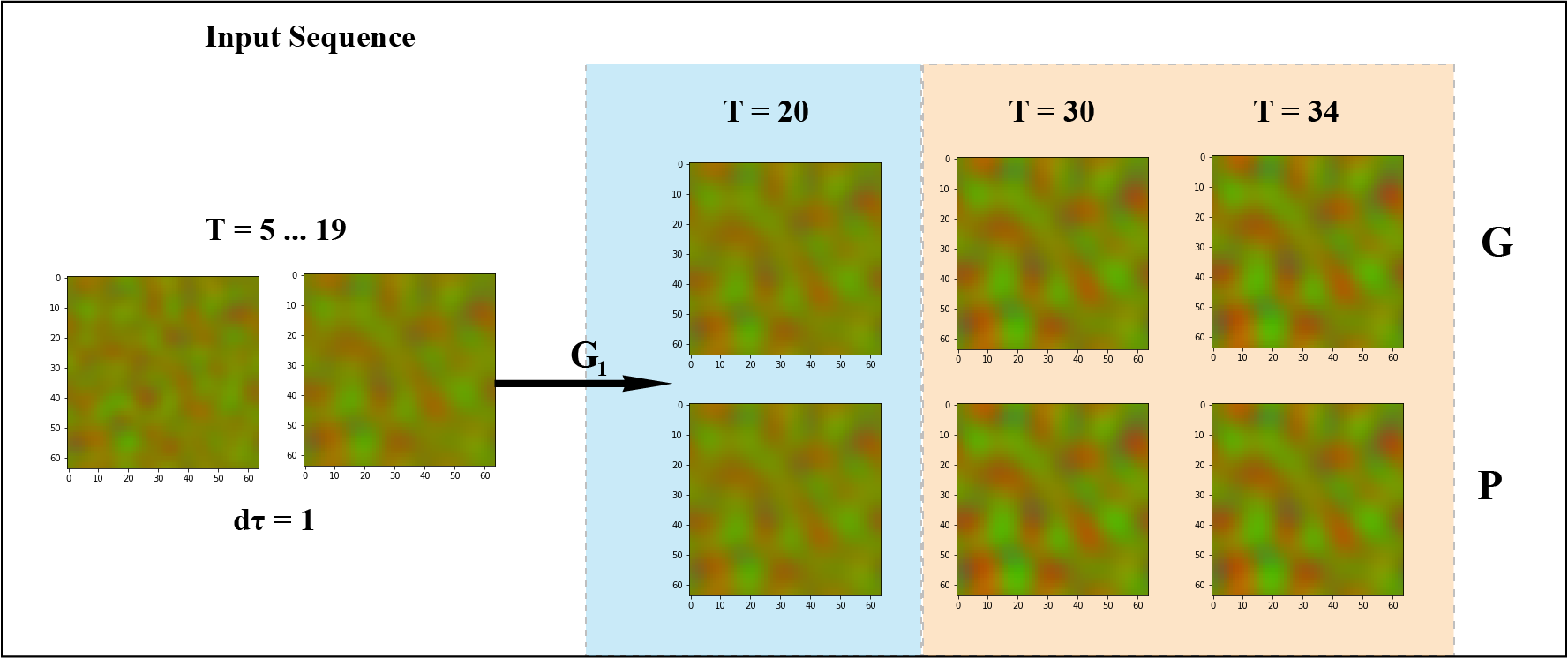}
	\caption{The sequences for the time-steps 5 to 20 of the evolution given as input to the model trained for $\mathcal{L}_1$ (Left) on a domain size of 64 by 64. The Blue region (Centre) corresponds to stages of evolution the model has seen via the training data and the Red region (Right) corresponds to stages of evolution the model hasn't seen. Clearly, the model is able to predict from an intermediate time-step as well (Interpolation Regime). G: Ground Truth, P: Predictions by the models.}
	\label{fig:Fig5}
\end{figure}

\begin{figure}
	\centering
	\includegraphics[width=14cm]{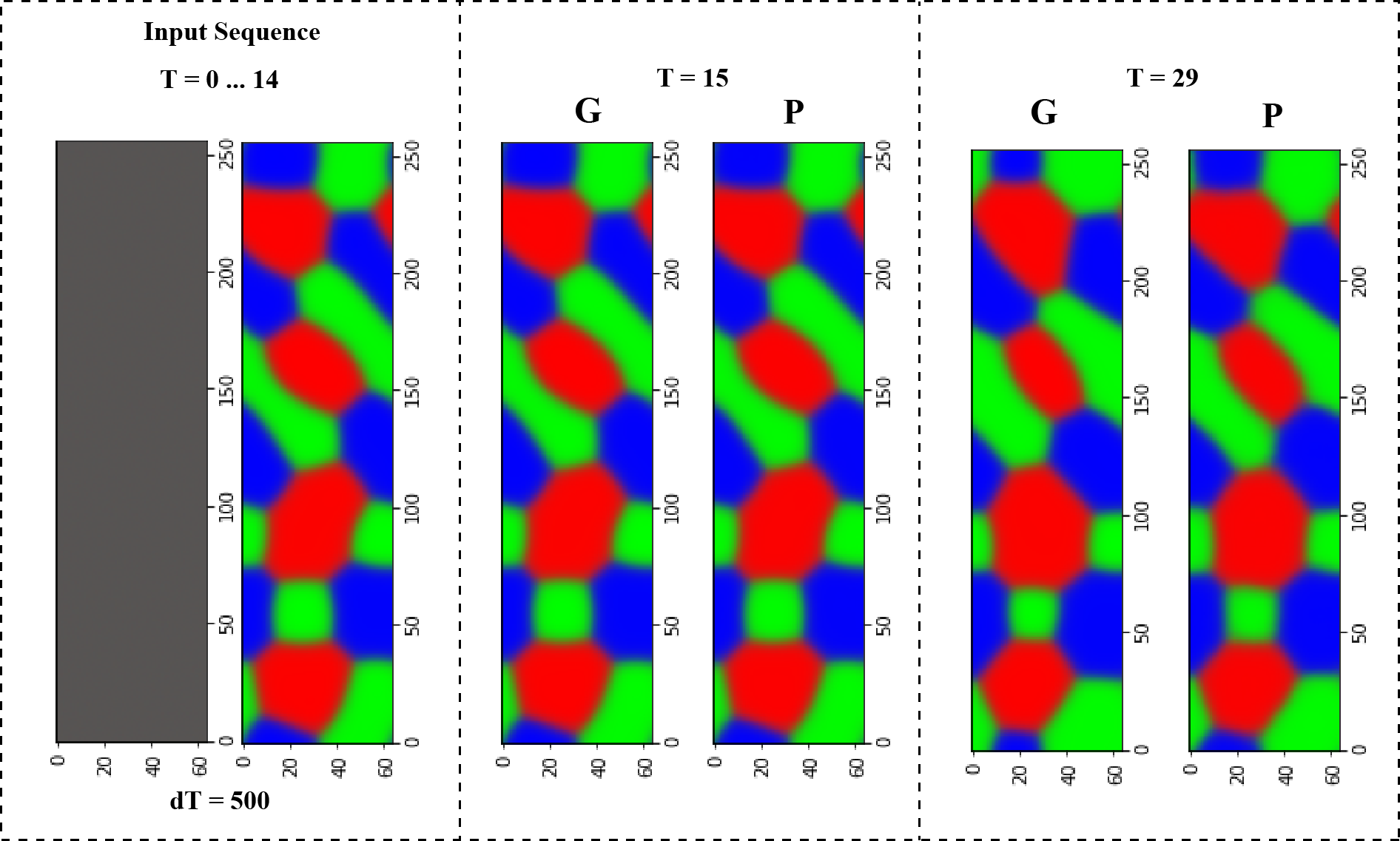}
	\caption{The sequences for the first 15 time-steps ($\mathcal{G}_{500}$) for a rectangular domain of size 64 by 256 are given as input (Left) to the model trained on a square domain of size 64 by 64. Clearly, the model is capable of predicting the solution on a mesh with a different geometry without incurring any additional errors as well. Hence, the proposed model is capable of learning a \textbf{mesh independent solution} to the problem. G: Ground Truth, P: Predictions by the model.}
	\label{fig:Fig7}
\end{figure}

To assess the scalability of the model, predictions were made on a grid size of 256 by 256, using the model that was trained on data corresponding to a 64 by 64 grid, which have been shown in Fig \ref{fig:Fig6}. As can be seen from Figures \ref{fig:Fig4} and \ref{fig:Fig6}, the model is capable of making accurate predictions on a larger mesh size without retraining. Furthermore, the transfer-ability of the trained model on a mesh with a different geometry was also tested. Figure \ref{fig:Fig7} shows the results of applying the models trained on a square mesh of dimensions 64x64 to make predictions on a rectangular mesh of dimensions 64x256. Clearly, the model is invariant of the mesh used to discretize the domain as well. 
\subsection{Computational Complexity}
\begin{table}
	\caption{Computation time for difference grid sizes. The Numerical solver was run for 20,000 steps for 5 different initial conditions, while the conv-LSTM GAN models recursively made predictions on a single K80 GPU. All times have been reported in seconds.}
	\centering
	\begin{tabular}{lccc}
    \hline \multicolumn{1}{c}{ Model Configuration } & 64x64 grid & 256x256 grid \\
    \hline Numerical Solver & $36.168$ & $909.022$ \\
    \hline $\mathcal{G}_{1}$ & $144.462$ & $828.836$ \\
    \hline $\mathcal{G}_{20}$ & $9.501$ & $74.177$ \\
    \hline $\mathcal{G}_{40}$ & $4.973$ & $40.319$ \\
    \hline $\mathcal{G}_{500}$ & $\mathbf{0.368}$ & $\mathbf{2.857}$ \\
    \hline
    \end{tabular}
	\label{tab:table2}
\end{table}
Another aspect to analyze is the computation times needed for making predictions using the proposed conv-LSTM GAN model, which have been tabulated in Table \ref{tab:table2}. Furthermore, another $0.742$ seconds are needed to load a trained model in memory onto the K80 GPU before predictions can be made, which is very small in comparison to the times mentioned in Table 2. As can be seen from the computation times, although the model offers only a small speedup for the case when $d\tau=1$, by selecting a larger $d\tau$ and utilizing the mesh-independence and generalization capabilities of the trained models, it is potentially possible to speed up predictions for forward and inverse problems by upto 3 orders of magnitude while incurring an acceptable amount of error in the solution. 

\subsection{Effect of increasing time-difference between subsequent inputs}
\begin{table}
	\caption{Results of predictions made by the trained conv-LSTM GAN model for a $d\tau$ value of 500.}
	\centering
	\begin{tabular}{lcccccc}
    \hline \multicolumn{1}{c}{ Model Configuration } & Metric & T=15 & T=29 & T=30 & T=44 \\
    \hline $d\tau=500, d_x=d_y=64$ & MSE & $1.24\times10^{-4}$ & $1.28\times10^{-3}$ & $1.76\times10^{-3}$ & $3.64\times10^{-3}$ \\
    & SSIM & $0.99870$ & $0.98956$ & $0.98697$ & $0.97367$ \\
    \hline
    \end{tabular}
	\label{tab:table_T500}
\end{table}
As can be inferred from Table \ref{tab:table1}, in general on increasing the $d\tau$ of the model, performance appears to degrade. Although the predictions made by the model are visually similar to the ground truth, since the metrics applied are pixel-based, the errors in prediction accumulate with time. Furthermore, since the data used to train $\mathcal{G}_{500}$ has a very high $d\tau$, the sequence reaches the near-equilibrium solution around T=30, therefore there is a very little increase in the error after this time-step, as can be seen in Table \ref{tab:table_T500}. However, as long as the time difference is kept within a controlled bound, the models allow one to effectively increase the time-step of the numerical solver and speed up the simulation, albeit while introducing an acceptable amount of error into the system. 

\section{Conclusion}
In summary, we show that U-net GAN models can effectively learn differential operators used in the coupled system of fourth-order Cahn-Hilliard equations through the physics-based simulation data, all while being completely agnostic to the underlying PDEs. Further, the use of conv-LSTM layers allows our model to capture the temporal nature of the data. We note that this model requires far lesser data and time to train compared to conventional surrogate models. In other words, our model can achieve the same level of accuracy as the conventional data-driven models without using large training datasets.


Furthermore, the model is also capable of speeding-up both forward and inverse calculations of the underlying PDEs by upto 3 orders of magnitude, all while introducing only a small amount of error into the system. A striking aspect of our model is that it is accurate even when predicting on larger meshes, making it mesh and scale-independent. This makes it a viable candidate to be used when scale-bridging during multi-scale modelling, with simulations performed at much larger scales.

Although the proposed U-net GAN models appear quite promising, there still exist issues with regards to the fall in accuracy on increasing the time-step of the data used for training, which can blow-up over time leading to a very different solution than what is expected. Another concern is with respect to the sensitivity of the model to the initial conditions. With the success of the Neural Operator on learning solutions to parametric PDEs accurately, we expect a future direction for the development of Neural-PDE solvers could involve the application of neural operators, such as the Fourier Neural Operator \cite{li2021fourier, wen2021ufno}, in the proposed U-net GAN framework and their use in tackling more challenging Phase-Field problems with $>2$ order parameters and for a wider range of initial and boundary conditions.

\bibliographystyle{unsrt}  
\bibliography{references}

\end{document}